\documentclass[twocolumn,showpacs,pre,aps,superscriptaddress]{revtex4}
\usepackage[latin1]{inputenc}
\usepackage{bm}
\usepackage{multirow}
\usepackage{amssymb}
\usepackage{amsbsy}
\usepackage{amsmath}
\usepackage{stmaryrd}
\usepackage{graphicx}
\usepackage{epsfig}

\begin{document}

\title{Occurrence of
exponential relaxation in closed quantum systems}

\author{Christian Bartsch}

\email{cbartsch@uos.de}

\affiliation{Fachbereich Physik, Universit\"at Osnabr\"uck,
             Barbarastrasse 7, D-49069 Osnabr\"uck, Germany}

\author{Robin Steinigeweg}

\email{rsteinig@uos.de}

\affiliation{Fachbereich Physik, Universit\"at Osnabr\"uck,
             Barbarastrasse 7, D-49069 Osnabr\"uck, Germany}
   
\author{Jochen Gemmer}

\email{jgemmer@uos.de}

\affiliation{Fachbereich Physik, Universit\"at Osnabr\"uck,
             Barbarastrasse 7, D-49069 Osnabr\"uck, Germany}

\date{\today}

\begin{abstract}
We investigate the occurrence of exponential relaxation in
a certain class of closed, finite systems on the basis
of a time-convolutionless (TCL) projection operator expansion
for a specific class of initial states with vanishing inhomogeneity.
 It turns out that exponential behavior is to be expected only if the
 leading order predicts the standard separation of timescales and if,
 furthermore, all higher orders remain negligible for the full
 relaxation time. The latter, however, is shown
 to depend not only on the perturbation (interaction) strength, but
 also crucially on the structure of the perturbation matrix. It is
 shown that perturbations yielding exponential relaxation have to
 fulfill certain criteria, one of which relates to the so-called 
 ``Van Hove structure''. All our results are verified by
the numerical integration of the full time-dependent
Schr\"odinger equation.
\end{abstract}


\maketitle

%
%

\section{Introduction} \label{sec-introduction}

A substantial part of linear non-equilibrium thermodynamics essentially
relies on a description by means of rate equations, often in the form
of master equations \cite{kubo1991}. The crucial quantities, such as the probability
to find the system in some state $i,j$ or the amount of particles,
energy, etc. at points $i,j$ in some space, are routinely
believed to follow equations like 
\begin{equation}
\frac{\partial}{\partial t} \, P_{i} = \sum_{j} R(j \rightarrow i)
\, P_{j} - \sum_{j} R(i \rightarrow j) \, P_{i}, \label{eq-rate}
\end{equation}
with time-independent transition rates from $i$ to $j$, $R(i
\rightarrow j)$. Pertinent examples are the decay of excitations in 
atoms, nuclear decay, etc. But also diffusive transport
phenomena belong to that class, since the diffusion equation can also be formulated to
take the above form (random walk dynamics). Another implementation of
that scheme is the (linear)
Boltzmann equation \cite{kubo1991,cercignani1988} where particle scattering is taken into account by
means of transition rates, and many more could be named. 
\\
However, regardless of the incontestable success of such descriptions,
the strict derivation of rate equations from underlaying principles
often remains a problem. Typically, the descriptiveness by means of rate equations is taken for granted. Since those rate
equations yield an exponential decay towards equilibrium, the basic question may be formulated as: How can an exponential
decay of some observable be derived from the Schr\"odinger equation? 

On the basis of quantum mechanics the most popular approach to this
question is probably Fermi's Golden Rule \cite{cohen1993}. Despite the
undisputed descriptive success of this scheme, it is simply derived from first
order perturbation theory, e.g., its validity generally breaks
down on a timescale much shorter than the resulting relaxation time. Therefore it can 
hardly describe a complete decay into equilibrium. 
One of the few concrete, concise derivations of exponential decay is the
Weisskopf-Wigner theory for the relaxation of excitations in an atom
due to the coupling of the atom to a zero-temperature, broad-band
electromagnetic field \cite{scully1997}. However, this theory is hardly generalizable,
since it only applies if just one state is coupled to a multitude of
others, rather than many states coupled to many others, as is typically
the case.
\\
A more abstract, rather fundamental approach has been suggested by Van
Hove \cite{vanhove1955,vanhove1957}. 
It is based on (infinite) quantum systems having continuous state
densities and interactions which are described by smooth functions
rather than discrete matrices. However, a lot of the findings for
discrete systems in the paper at hand are quite parallel to Van
Hove's, as will be pointed out below.
\\ 
Other approaches are based on projection operator techniques, in particular
the well-known Nakajima-Zwanzig (NZ) method. This method is
commonly used in the context of open quantum systems, i.e., systems
that allow for a partition according to a considered system (or simply ``system'') and an environment \cite{kubo1991,weiss1999}. For a specific choice of the initial condition, as pointed out below, the
projection onto the system's degrees of freedom eventually leads
to an autonomous master equation describing the dynamics of the system, based on a
systematic perturbation expansion. But in general, due to the
complexity of higher orders, only the leading order is taken into
account. In the paper at hand we will demonstrate that this truncation
may produce wrong results even and especially for the case of fast
decaying correlation functions and arbitrarily weak interactions.
\\
A further approach to this topic is based on the description of quasi-particle dynamics in many-particle systems by the use of Green's functions \cite{kadanoff1962}.
These considerations indicate the validity of a Boltzmann equation.
\\
In the present paper we will employ another projection operator
technique, the so-called time-convolutionless (TCL) method \cite{fulinski1967,fulinski1968,hanggi1977,grabert1977,chaturvedi1979,breuer2007}.
In the following we will follow the TCL-method as detailed in \cite{breuer2007}.
In Sec.~\ref{sec-model} we introduce our rather abstract Hamiltonian for
a ``closed quantum'' system (consisting of an unperturbed part and a perturbation) and define an also rather
abstract observable, the dynamics of which we are going to
investigate. In Sec.~\ref{sec-tcl2} we demonstrate how the TCL
technique can be used to compute the above dynamics of the variable.
(This is somewhat reminiscent of projection techniques using
``correlated projectors'' \cite{blanga1996,budini2006}.) We tune our models such that a leading
order truncation predicts exponential decay. For a ``random
interaction'' this prediction turns out to be correct, as is verified by the numerically exact
solution of the full time-dependent Schr\"odinger equation. In the
following Sec.~\ref{sec-tcl4} non-random (``structured'') perturbation
matrices are
discussed in more detail. While a leading order truncation still
predicts exponential relaxation, it is demonstrated that this
prediction may fail even for arbitrarily large models and arbitrarily
small interactions. This breakdown stems from the fact that higher order contributions are not
negligible if the interaction matrix violates certain criteria.
Before we will close with a summary and conclusion in
Sec.~\ref{sec-summary}, these criteria will be also related to
those conditions which Van Hove postulated in order to explain the
occurrence of exponential relaxation.

%
%

\section{Models, Observables and Interpretation of Dynamics} \label{sec-model}
In the present paper we will analyze quantum models which are much
simpler than most of  the examples mentioned in the introduction. They are
defined on a very general, rather formal level and are not meant to
describe any specific, realistic quantum system in great detail. The 
Hamiltonian is taken to consist of a local part $H_{0}$ and an interaction
part $V$ such that $H = H_{0} + V$. In particular, $V$ is assumed
to take the special form of an ``off-diagonal block structure'' in
the eigenbasis of $H_{0}$, that is, the matrix representation of
$H$ may be written as
\begin{equation}
\centering
H = \left(%
 \begin{array}{ccccc|ccccc}
 \ddots & &                                  & & 0      &        & &                                  & &       \\
        & &                                  & &        &        & &                                  & &       \\
        & & \frac{i}{n-1} \, \delta \epsilon & &        &        & & v                                & &       \\
        & &                                  & &        &        & &                                  & &       \\
 0      & &                                  & & \ddots &        & &                                  & &       \\
 \hline
        & &                                  & &        & \ddots & &                                  & & 0     \\
        & &                                  & &        &        & &                                  & &       \\
        & & v^{\dagger}                      & &        &        & & \frac{j}{n-1} \, \delta \epsilon & &       \\
        & &                                  & &        &        & &                                  & &       \\
        & &                                  & &        & 0      & &                                  & & \ddots\\
 \end{array}
 \right)%
 \label{eq-hamiltonian}
 \end{equation}
or, equivalently to the above notation, $H=H_0+V$ may also be written as
\begin{eqnarray}
H_0 &=& \sum_{i = 0}^{n-1} \frac{i}{n-1} \, \delta \epsilon \; | i
\rangle \langle i | + \sum_{j = 0}^{n-1} \frac{j}{n-1} \, \delta
\epsilon \; | j \rangle \langle j | \ , \nonumber \\
V&=& \left( \sum_{i,j = 0}^{n-1} v_{ij} \; | i \rangle \langle j | +
\text{H.c.} \right),
\end{eqnarray}
where $| i \rangle,| j \rangle$ form the basis in which
(\ref{eq-hamiltonian}) is represented.
Obviously, the complete Hilbert space is divided into two
subspaces, where $i$ runs through the states of the first and $j$
through the states of the second subspace,
respectively. Obviously, $H_0$ may correspondingly be separated
into two parts which we only specify very roughly at this point by two
parameters: There are two identical ``bands'' with width $\delta \epsilon$ and $n$
equidistant energy levels each.
\\
The average strength of the interaction $V$ is measured by
\begin{equation}
\lambda^2 = \frac{1}{n^2} \sum_{i,j = 0}^{n-1} | v_{ij} |^{2} \; .
\end{equation}
In our first example in Sec. \ref{sec-tcl2} we take the matrix elements $v_{ij}$ in
the off-diagonal blocks to be Gaussian, complex, random numbers. For
the other examples $V$ will be specified below. In all cases the matrix elements of $V$ in the
diagonal blocks are all zero, just to keep the picture as simple as possible.
\\
We will investigate the (relaxation) dynamics of an abstract
observable $a$, represented by an operator $A$, which is
chosen in such a way that
\begin{equation}
[ A, H_{0} ] = 0 \, , \; \text{Tr}\{ A \} = 0 \, , \; \text{Tr}\{
A^{2} \} = 1 \, .
\end{equation}
The first of these properties states that $A$ is diagonal in the
eigenbasis of $H_{0}$, while the remaining two properties do not
mean crucial restrictions on $A$. While all of the following will be
correct for any $A$ featuring the above properties, we mainly
concentrate in our examples on  ``binary'' operators, i.e., operators
featuring only two different eigenvalues, namely, $+\, 1 / \sqrt{2 \,
  n}$ in one subspace and $-\, 1 / \sqrt{2 \,
  n}$ in the other. This means that $a\equiv \text{Tr}\{ A \rho\}=+\, 1
/ \sqrt{2 \,n}$ indicates that the system entirely occupies one
subspace and $a=-\, 1 / \sqrt{2 \,n}$ indicates that it entirely
occupies the other subspace. (Here, $\rho$ is the density matrix for the
state of the system.) If and only if  $a(t)$ is found to relax exponentially
to zero, the system allows for a merely statistical interpretation
entirely beyond quantum physics: it is then in accord with a system
featuring two distinguishable states in between it can ``hop'' with a
given transition rate, the latter being equal for both directions. $a$
then represents the difference between the probabilities of finding it in
one or the other state, respectively.
\\
In an abstract way the above model 
may represent many physical situations.
It may be viewed as a simplified model for the exchange of an
excitation between, e.g., two weakly coupled atoms, molecules, quantum
dots, etc. $A$ then represents the probability to find atom $1$ excited, subtracted by the probability
to find atom $2$ excited, $V$ represents the coupling in this scenario. Or it may model the momentum dynamics of a particle bound
to one dimension which possibly changes its direction (forward-backward) due to some
scattering. In a many-particle system the current operator could be identified with $A$ and $V$ may stand for a particle-particle interaction.
This way the dynamics of the current autocorrelation function could be investigated based on the framework below.
More detailed information about such models can be found
in \cite{michel2005,gemmer2006,steinigeweg2006,breuer2006,steinigeweg2007,kadiroglu2007}.

%
%

\section{TCL scheme and choice of the projection operator} \label{sec-tcl}

In this section we give a short overview of the
time-convolutionless (TCL) projection operator technique
\cite{chaturvedi1979,breuer2007}. Furthermore, we introduce the
pertinent equations which are applied to models with various 
interactions in Sec. \ref{sec-tcl2} and Sec. \ref{sec-tcl4}. 
A detailed derivation of these
equations is beyond the scope of this paper and can
be found in \cite{breuer2006,breuer2007}.
\\
The TCL method is a projection operator technique such as the
well-known Nakajima-Zwanzig technique
\cite{nakajima1958,zwanzig1960}. Both are applied in order to
describe the reduced dynamics of a quantum system with a
Hamiltonian of the type $H = H_{0} + V$. Generally, the full
dynamics of the system are given by the Liouville-von Neumann
equation,
\begin{equation}
\frac{\partial}{\partial t} \, \rho(t) = -\imath \, [ \, V(t),
\rho(t) \, ] = \mathcal{L}(t) \, \rho(t) \, . \label{eq-liouville}
\end{equation}
(Now and in the following all equations are denoted in the
interaction picture.) In order to describe the reduced dynamics of
the system, one has to construct a suitable projection operator
$\mathcal{P}$ which projects onto the relevant part of the density
matrix $\rho(t)$. $\mathcal{P}$ has to satisfy the property
$\mathcal{P}^{2} \,\rho(t) = \mathcal{P} \,\rho(t)$.
Recall that in our case the relevant variable is
chosen as the expectation value $a(t)$ of the binary operator $A$.
For initial states $\rho(0)$ with 
\begin{equation}
\mathcal{P} \, \rho(0) = \rho(0)
\label{init}
\end{equation}
the TCL method yields a closed
time-local equation for the dynamics of $\mathcal{P} \, \rho(t)$,
\begin{equation}
\frac{\partial}{\partial t} \, \mathcal{P} \, \rho(t) =
\mathcal{K}(t) \, \mathcal{P} \, \rho(t) \label{eq-tcl2-1}
\end{equation}
with
\begin{equation}
\mathcal{K}(t) = \sum_{i=1}^{\infty} \mathcal{K}_{i}(t) \, .
\label{eq-tcl2-2}
\end{equation}
The TCL technique avoids the usually troublesome time
convolution which appears, e.g., in the context of the
Nakajima-Zwanzig technique.
Eq.~(\ref{eq-tcl2-1}) and (\ref{eq-tcl2-2}) represent a formally exact
perturbative expansion.\\
A brief comment on initial conditions should be made here.
If (\ref{init}) is not fulfilled, of course an additional inhomogeneity appears on 
the r.h.s. of (\ref{eq-tcl2-1}). This may change the solutions of (\ref{eq-tcl2-1}) drastically, 
c.f. \cite{romero2004} and references therein. However, for the model to be addressed 
below, there is substantial numerical evidence that, for a large set of 
initial states that do not fulfill (\ref{init}), the dynamics are nevertheless 
reasonably well described by (\ref{eq-tcl2-1}) (without inhomogeneity) 
\cite{esposito2003,gemmereur2006,gemmereurlett2006,breuer2006,michel2005,gemmer2004}. 
Having mentioned this issue we consider in the following exclusively 
initial states in accord with (\ref{init}).
\\
For many models the odd
cumulants of the expansion (\ref{eq-tcl2-2}) vanish: $\mathcal{K}_{2i+1}(t) = 0$. This
will turn out to apply to our model as well.
The lowest non-vanishing order scales quadratically with $\lambda$
and reads
\begin{equation}
\mathcal{K}_{2}(t) = \int_{0}^{t} dt_1 \, \mathcal{P} \,
\mathcal{L}(t) \, \mathcal{L}(t_1) \, \mathcal{P} \; .
\end{equation}
For the fourth order term one finds
\begin{eqnarray}
\mathcal{K}_4(t) &=& \int_{0}^{t} \! dt_1 \int_{0}^{t_1} \! dt_2 \int_{0}^{t_2} \! dt_3 \nonumber \\
&& \mathcal{P} \, \mathcal{L}(t) \, \mathcal{L}(t_1) \, \mathcal{L}(t_2) \, \mathcal{L}(t_3) \, \mathcal{P} \nonumber\\
&-& \mathcal{P} \, \mathcal{L}(t) \, \mathcal{L}(t_1) \, \mathcal{P} \, \mathcal{L}(t_2) \, \mathcal{L}(t_3) \, \mathcal{P} \nonumber \\
&-& \mathcal{P} \, \mathcal{L}(t) \, \mathcal{L}(t_2) \, \mathcal{P} \, \mathcal{L}(t_1) \, \mathcal{L}(t_3) \, \mathcal{P} \nonumber \\
&-& \mathcal{P} \, \mathcal{L}(t) \, \mathcal{L}(t_3) \, \mathcal{P} \, \mathcal{L}(t_1) \, \mathcal{L}(t_2) \, \mathcal{P} \; . %
\label{eq-tcl4-1}
\end{eqnarray}
\\
Note that the TCL approach is commonly used in the context of open
quantum systems
\cite{breuer2007,nakajima1958,zwanzig1960,kubo1991}. The TCL method
is, however, also applicable to our
closed quantum system.
\\
To those ends, we define the projection operator ${\cal P}$ by
\begin{equation}
\mathcal{P} \, \rho(t) \equiv \frac{1}{2n} \, \hat{1} + A \, \text{Tr} \{ A \, \rho(t) \} =
\frac{1}{2n} \, \hat{1} + A \, a(t) \; . \label{eq-proj}
\end{equation}
As already mentioned above, $\cal P$ is constructed to project
onto the time-dependent expectation value $a(t)$ of the binary
operator $A$, in the Schr\"odinger picture. But since $A$ commutes
with $H_{0}$, this expectation value is identical in the
interaction and the Schr\"odinger picture. The full dynamics
[Hilbert space: dimension $2n$,  Liouville space of density
matrices: dimension $(2n)^{2}$] is broken down to the time
evolution of the single variable $a(t)$, all other information is
neglected. As a suitable initial condition we can then choose
$\rho(0) = (1 / 2n) \, \hat{1} + (1 / \sqrt{2n}) \, A$ which
implies $a(0)= 1 / \sqrt{2n}$.
Inserting Eq.~(\ref{eq-proj}) into
Eq.~(\ref{eq-tcl2-1}) yields the closed equation
\begin{equation}
\dot{a}(t) = \sum_{i=1}^{\infty} K_{i}(t) \, a(t) \label{eq-adyn}
\end{equation}
with $K_{i}(t) = \text{Tr} \{ A \, \mathcal{K}_{i}(t) \, A \}$.
Due to Eq.~(\ref{eq-proj}), the second order term reads
\begin{equation}
K_{2}(t) = -\int_{0}^{t} d t' \, C(t') \label{eq-tcl2-3} \; ,
\end{equation}
where the two-point correlation function $C(t')$ is given by
\begin{equation}
\label{corrf}
C(t') = \text{Tr} \Big \{ \imath [ \, V(t), A \, ] \; \imath [ \,
V(t_{1}), A \, ] \Big \} \; , \quad t' \equiv t - t_1 \; .
\end{equation}
A rather lengthy but straightforward calculation yields for the fourth order
\begin{eqnarray}
K_{4}(t) &=& \int_{0}^{t} \! dt_1 \int_{0}^{t_1} \! dt_2\int_{0}^{t_2} \! dt_3 \ I_{1}+I_{2}+I_{3}+I_{4} , \nonumber \\
I_{1}=&& \! \text{Tr} \Big \{ [ \, V(t_1), [V(t), A] \, ] \, [ \, V(t_2), [V(t_3), A] \, ] \Big \} , \nonumber \\
I_{2}=&-& \! C(t-t_1) \; C(t_2-t_3) \ , \nonumber \\
I_{3}=&-& \! C(t-t_2) \; C(t_1-t_3) \ , \nonumber \\
I_{4}=&-& \! C(t-t_3) \; C(t_1-t_2) \; . \label{eq-tcl4-2}
\end{eqnarray}

\section{Second order TCL and completely random interaction} \label{sec-tcl2}

\begin{figure}[htb]
\centering
\includegraphics[width=7cm]{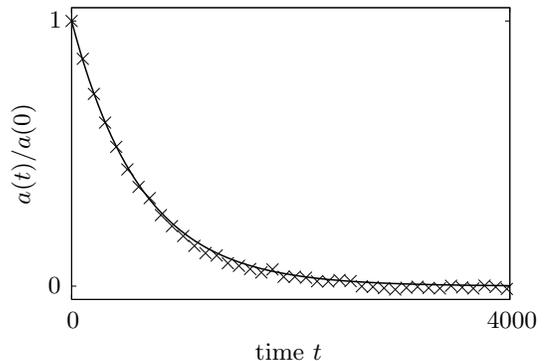}
\caption{Time evolution of the expectation value $a(t)$ for an
interaction with completely random $v_{ij}$. The numerical result
(crosses) indicates exponential behavior and is in very good
agreement with the theoretical prediction (\ref{eq-astat}) of
second order TCL (continuous curve). The system parameters $n =
1000$, $\delta \epsilon = 0.5$, $\lambda = 2.5 \cdot 10^{-4}$
fulfill the conditions (\ref{eq-crit}).} \label{fig-randWW}
\end{figure}

In this section we apply the equations in second order TCL to a model with the completely random interaction
introduced in Sec. \ref{sec-model}.
The function $C(t')$ in Eq.~(\ref{corrf}) is identical to the autocorrelation function of the
interaction, since it can also be written as
\begin{equation}
\label{cosini}
C(t') = \frac{4}{n} \sum_{i,j = 0}^{n-1} |v_{ij}|^2 \, \cos [ \,
\omega_{ij} \, (t - t_1) \, ]
\end{equation}
with frequencies $\omega_{ij} = (i \! - \! j) / (n \! - \! 1) \,
\delta \epsilon$ corresponding to $H_0$. Here, just like in many other examples, $C(t')$ decays within the correlation time $\tau_C$
which is of the order of $\tau_C \approx 4 \pi / \delta \epsilon$
for our model. Afterwards the integral $K_2(t)$ becomes
approximately time-independent and assumes a constant value $R$ until
the ``Heisenberg time'' $T = 2 \pi \, n / \delta \epsilon$ is reached. This
behavior can be inferred from integrating (\ref{cosini}) and
exploiting the properties of the sinc-function. From this
analysis also $R$ may be found with an accuracy determined by the law
of large numbers. Thus the second order approximation of Eq.~(\ref{eq-adyn}) eventually
results in
\begin{equation}
\dot a(t) = -R \; a(t) \, , \quad R \approx \frac{4 \pi \, n \,
\lambda^{2}}{\delta \epsilon} \; . \label{eq-astat}
\end{equation}
We hence obtain a rate equation featuring the form of 
Eq.~(\ref{eq-rate}) and thus exponential dynamics for $a(t)$. The
solutions for $a(t)$ decay exponentially with a relaxation time $\tau_R = 1/ R$. However, this result is only valid
within the boundaries $\tau_C \ll \tau_R < T$, because $K_2(t)$
can only be considered as time-independent up to the Heisenberg
time. Recall to this end that
our model features equidistant energies such that $C(t')$ is
strictly periodic with $T$. These two boundaries also result in
two necessary criteria for the system parameters which have to be
fulfilled in order to produce the occurrence of exponential dynamics,
\begin{equation}
\frac{16 \pi^{2} \, n \, \lambda^{2}}{\delta \epsilon^{2}} \ll 1
\, , \quad \frac{8 \pi^{2} \, n^{2} \, \lambda^{2}}{\delta
\epsilon^{2}} > 1 \, . \label{eq-crit}
\end{equation}
Remarkably, the whole derivation of the rate equation using second order TCL does not depend on the details of the
interaction, i.e., the individual absolute values of the single matrix
elements as well as their relative phases are not relevant. We should already
mention here that the ``structure'', which we are going to introduce into
the interaction in the following section, only concerns those details,
hence the second order contribution $K_2$ will be the same in all our
following examples.
\\
In Fig.~\ref{fig-randWW} the numerical
solution of the Schr\"odinger equation is shown for the above
repeatedly mentioned random interaction and compared with the TCL
prediction. All parameters (the width of the Gaussian distribution according to which the matrix elements of $V$
are generated, the bandwith, etc.)  are adjusted such that the
criteria (\ref{eq-crit}) are well satisfied. This solution is
obtained by exact diagonalization. And in fact, we find a very
good agreement with the theoretical prediction of second order
TCL.

%
%

\section{Fourth Order TCL and Non-Random Interactions} \label{sec-tcl4}

In this section we will be concerned with the structure
of the interaction matrix and, especially, its influence on the time
evolution of the expectation value $a(t)$. It will be demonstrated
that the theoretical prediction (\ref{eq-astat}) of second order
TCL fails to describe the numerically exact solution of the
Schr\"odinger equation correctly for certain ``interaction types'',
even and especially if the conditions (\ref{eq-crit}) are fulfilled, i.e., the ``strength'' is ``adequate''. We will outline that this
failure stems from the fact that the fourth order
contribution of the TCL expansion is not negligible on the
relaxation timescale which is obtained from second order
TCL. 
However, the exact evaluation of
$K_4(t)$ turns out to be almost impossible, analytically
and numerically. Instead we will present feasible estimations of
$K_4(t) / K_2(t)$ based on suitable approximations of $K_4(t)$ called $S(t)$ (see (\ref{vierzwei}, \ref{nae})). Whenever
\begin{equation}
q(t) \equiv \frac{S(t)}{K_{2}(t)} < 1
\label{k4nae}
\end{equation}
is violated the influence of higher order terms is not negligible.
If this is the case for times $t$ of the order of or shorter than
$\tau_R$, no exponential relaxation will result.

\subsection{Uniform Interactions and Van Hove structure}

\begin{figure}[htb]
\centering
\includegraphics[width=7cm]{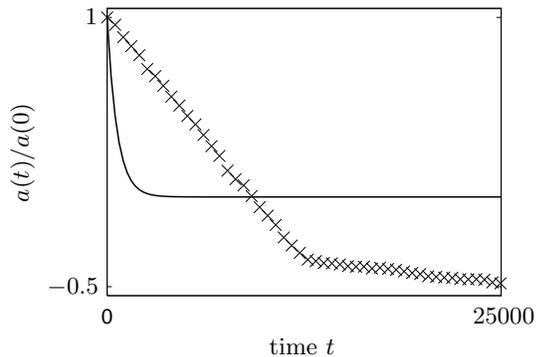}
\caption{Time evolution of the expectation value $a(t)$ for an
interaction with $v_{ij} = \lambda$. The theoretical prediction
(\ref{eq-astat}) of second order TCL (continuous curve) fails to
describe the numerical solution (crosses) correctly, although the
system parameters $n = 1000$, $\delta \epsilon = 0.5$, $\lambda =
2.5 \cdot 10^{-4}$ still fulfill the conditions (\ref{eq-crit}).
$V$ violates the Van Hove structure.}
\label{fig-constWW}
\end{figure}

\begin{figure}[htb]
\centering
\includegraphics[width=5cm]{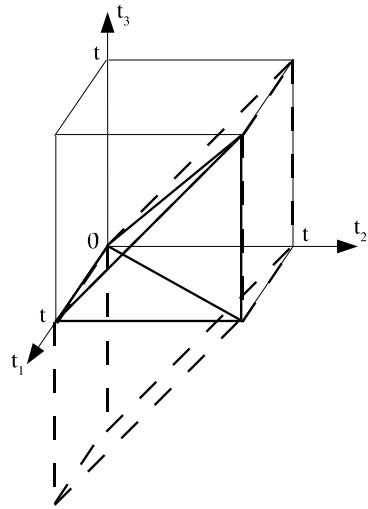}
\caption{Sketch for the integration volume of $K_{4}(t)$ for fixed
$t$. The cube with the edge length $t$ is drawn with thin lines,
the actual integration volume is marked with thick lines. The
dashed lines represent the changed integration volume which is
used in the approximations.} \label{fig-cube}
\end{figure}

Let us start with an example. Fig.~\ref{fig-constWW} shows the
time evolution of the expectation value $a(t)$ for an interaction
with $v_{ij} = \lambda$.  This type of interaction is, of course, highly
non-random, since all matrix elements have the same absolute value and
phase. The second order approximation obviously yields a wrong
description for this interaction structure, that is, the dynamics
are not exponential, although both of the conditions
(\ref{eq-crit}) are well fulfilled. It should be remarked again
that the observed non-exponential behavior definitely is a
structural issue. For instance, it can not be ``repaired'' by simply
decreasing the overall interaction strength, because this decrease
would eventually lead to the violation of the criteria
(\ref{eq-crit}).
\\
To analyze this model we now develop our first estimate $S(t)$ for $K_{4}(t)$
which concerns the timescale $t > \approx \tau_{C}$.
We start from from Eq.~(\ref{eq-tcl4-2}), where we abbreviate the
triple time integration by a single ``$\int$''. One may hence write
$K_{4}(t) = \int I_1 + I_2 + I_3 + I_4$.
 Fig.~\ref{fig-cube} shows
a sketch for the integration volume of $K_{4}(t)$ in the
$3$-dimensional space which is spanned by $t_1$, $t_2$, $t_3$. The
integration does not run over the whole cube with the edge length
$t$, but only over the region where $t_3 \leq t_2 \leq t_1$ holds.
\\
$C(t-t_1)$ is the autocorrelation function of the interaction
which has already been mentioned in (\ref{corrf}). Recall that
$C(t')$ is only different from zero around $t' = 0$ in a small
interval of the width $\tau_C$. Thus, the integrands $I_2$,
$I_3$, $I_4$ are only different from zero in a small volume around
the region where both of the arguments are equal for each of the
two multiplied correlation functions.
\\
First of all let us focus on $I_{3}$ as well as $I_{4}$. The
integrand $I_{3}$ contributes to $K_4(t)$ for $t = t_2$ and
$t_1=t_3$, while the integrand $I_{4}$ contributes to $K_4(t)$ for
$t = t_3$ and $t_1 = t_2$, respectively. 
The sketch in Fig.~\ref{fig-cube} displays that 
both of these regions overlap
only in the vicinity of one single point with the integration
volume of $K_{4}(t)$, namely, at the point where all arguments are
equal to $t$. Especially, this overlap does not increase with $t$.
Therefore the triple time integration is estimated by $\int I_{3}
\approx \int I_{4} \approx C(0)^2 \, \tau_C^3$. Using the estimate $R
\approx C(0) \, \tau_C$, we eventually obtain for the ratio between the
contributions from $ I_{3} , I_{4}$ to the fourth order and the second
order $K_2(t>\tau_C)$ for times $t>\tau_C$
\begin{equation}
\frac{\int I_3}{R} \approx \frac{\int I_4}{R} \approx C(0) \,
\tau_C^2 \approx \frac{\tau_C}{\tau_R} \equiv \alpha \; .
\end{equation}
Recall that the derivation of exponential behavior within second
order TCL has required $\tau_C \ll \tau_R$ or, equivalently,
$\alpha \ll 1$ such that the contributions to $K_4(t)$ which arise
from $I_3$ and $I_4$ are negligible, at least in comparison with
$R$.
\\
Analogous conclusions cannot be made for the term $I_2$, because
its overlap with the integration volume is larger and grows with
$t$. We have to find another estimation for the contributions of
$I_2$ as well as $I_1$, of course. Our estimation is based on the fact that neither
$I_2$ nor $I_1$ can decay on a shorter timescale than $\tau_{C}$
in any possible direction of the $(t_1,t_2,t_3)$-space. This fact
is obviously correct for $I_{2}$. But what about $I_1$? Since the
term $I_1$ consists of summands which have the typical form
\begin{equation}
v_{ab} \, v_{bc} \, v_{cd} \, v_{da} \, \mathrm{e}^{-\imath \,
\omega_{ab} \, t} \, \mathrm{e}^{-\imath \, \omega_{bc} \, t_1} \,
\mathrm{e}^{-\imath \, \omega_{cd} \, t_2} \, \mathrm{e}^{-\imath
\, \omega_{da} \, t_3} \; ,
\end{equation}
only those frequencies and, especially, those largest
frequencies in $I_1$ which have already appeared in $C(t')$ contribute
significantly to $I_1$.
Consequently, $I_1$ can never decay faster than $C(t')$ in any
possible direction of the $(t_1,t_2,t_3)$-space. In the (possibly unrealistic) ``best
case'' $I_1+I_2$ decays within $\tau_C$ around the
point $I_{1/2}(t,t,t,t) = I_{1/2}(0,0,0,0) \equiv I_{1/2}(0)$. We
can therefore estimate the value of $q(\tau_C)$ by
\begin{equation}
\label{vierzwei}
\frac{\int I_1 + I_2}{R} \approx \frac{[ \, I_1(0) - C(0)^2 \, ]
\, \tau_{C}^3}{C(0) \, \tau_C} = \left [ \frac{I_1(0)}{C(0)^2} - 1
\right ] \alpha \equiv \beta \; .
\end{equation}
$\beta$ is a lower bound for the ratio between the fourth and the
second order of TCL for times $t>\tau_C$. If $\beta \approx 1$ or even larger, then
$K_4(t>\tau_C)$ dominates $K_2(t>\tau_C)$, that is, exponential behavior in
terms of the second order prediction cannot occur. But
$\beta \ll 1$, however, does not allow for a strict conclusion,
since a slower decay of $I_1$ or $I_2$, as the case may be, raises
their contribution to $K_4(t)$. Nevertheless, the condition $\beta
\ll 1$ is an additional criterion for the occurrence of exponential
decay which involves the structure of
$V$.
\\

In the following we will discuss why and to what extend $\beta$
and, especially, the ratio $I_1(0) / C(0)^2$ is related to the
conditions which have been postulated by Van Hove for the interaction $V$ in
order to explain the onset of exponential relaxation, see
\cite{vanhove1955,vanhove1957}. To this end, let us define a
hermitian operator $G$ by
\begin{equation}
G \equiv [ \, V, [ V, A ] \, ] \; .
\end{equation}
A straightforward calculation yields
\begin{equation}
I_1(0) = \text{Tr} \{ G^2 \} = \sum_{i, j} | G_{ij} |^2 \; , \quad
C(0) = \text{Tr} \{ A \, G \} \; ,
\end{equation}
where $G_{ij}$ represents the matrix elements of $G$ in the
eigenbasis of $H_{0}$. Furthermore, let us also introduce the
superoperator $\cal D$ which is given by
\begin{equation}
\mathcal{D} \, M \equiv \sum_i | i \rangle \, M_{ii} \, \langle i
|
\end{equation}
and projects any operator $M$ onto its diagonal elements in the
eigenbasis of $H_{0}$. Then the expression
\begin{equation}
( A, G ) \equiv \text{Tr} \{ A \, \mathcal{D} \, G \}
\end{equation}
defines an inner product between the operators $A$ and $G$,
because $( A, G ) = ( G, A )^*[=( G, A )]$ holds and $( A, A ) = 1$, as well as
$( G, G ) = \sum_i G_{ii}^{2}$, are both positive, real numbers. The Schwartz
inequality $( A, G )^2 \leq ( A, A )^2 \, ( G, G )^2$ can consequently
be formulated. By the use of $( A, G ) = C(0)$ we
eventually obtain
\begin{equation}
C(0)^2 \leq \sum_{i} G_{ii}^2 \leq \sum_{i, j} | G_{ij} |^2
=I_1(0) \; , \label{eq-schwartz}
\end{equation}
i.e., $I_1(0) / C(0)^2 \geq 1$. $C(0)^2$ is at most as large as
the sum of the squared diagonal elements of $G$, according to the
above equation. Therefore $I_1(0) / C(0)^2 \approx 1$ and hence
sufficiently small $\beta$ can  only be realized if the diagonal
elements of $G$ and thus the diagonal elements of $V^{2}$ are as
large as possible in comparison with the remaining non-diagonal
elements of $V^{2}$ ($G$). In principle, this is essentially what Van Hove proclaimed
\cite{vanhove1955,vanhove1957}.
\\
In this sense, we define the ``Van Hove structure'' in the context
of finite quantum systems: The interaction V is said to feature
Van Hove structure if
\begin{equation}
\label{vanh}
\beta' \equiv \frac{I_1(0)}{C(0)^2} \, \alpha \ll 1\ ,
\end{equation}
while all conditions of second order TCL are
simultaneously kept, of course. The latter refers
to the validity of Eq.~(\ref{eq-crit}). The comparison with
(\ref{vierzwei}) shows that the Van Hove structure implies
$\beta\ll 1$ and hence the relaxation may possibly be exponential, as
described by the second order. Since the evaluation of $\beta'$ is much
more efficient than the complete computation of fourth order TCL
(there is no time dependence left, e.g., $I_1(0)$ only
depends on $t = t_1 = t_2 = t_3 = 0$), the Van Hove structure
eventually is an assessable criterion for the possible occurrence of
exponential decay. 
It is a criterion in the sense that only if (\ref{vanh}) is satisfied, a use
of the second order approximation is justified for any time longer than the correlation time, i.e.,
$t>\tau_C$.
\\
Let us now apply these results to the already introduced models with random and
non-random $(v_{ij}=\lambda)$  interactions, respectively. The only term which varies
for the different models is $I_1(0)$, since 
the terms $C(0)^2\approx n^2\lambda^4$ and $\alpha \approx
16\pi^2n\lambda^2/\delta\epsilon^2$ (again with an accuracy set by the
law of large numbers for the random interaction) are the same for
random and non-random interactions. For the random interaction a
straightforward calculation leads to
\begin{equation}
I_1(0) = 32 \, n^2 \, \lambda^4 \, , \quad \beta' = 2 \, \alpha  \ll 1
\end{equation}
such that the random interaction indeed features Van
Hove structure. This agrees with
the numerical results in Fig.~\ref{fig-randWW} which yielded
exponential relaxation. In the case $v_{ij} = \lambda$, however,
we finally obtain
\begin{equation}
I_1(0) = 16 \, n^3 \, \lambda^4 \; , \quad \beta' = \frac{16 \pi^2
\, n^2 \, \lambda^2}{\delta \epsilon^2} \; ,
\label{vanhconst}
\end{equation}
where $\beta' > 1$, according to Eq.~(\ref{eq-crit}).
The absence
of the Van Hove structure already suffices to explain the breakdown of exponential
behavior in Fig.~\ref{fig-constWW}.
\\
One may nevertheless be inclined to argue that the Van Hove structure is not the
crucial difference between those two cases but simply the
randomness of the matrix elements (which possibly induces quantum chaos). We therefore present a
counter-example which immediately disproves such an argument. The
example is slightly different from the others, since the complete
system is not partitioned into equally large subspaces. $n_{1}$ and $n_{2}$ define the number of levels of the respective subspaces.
One subspace consists of only one state ($n_{1}=1$). Thus, in the matrix $V$
there is only a single column with non-zero elements and a single
row, respectively. Although these non-zero elements are chosen to
be all equal (non-random), it can be shown that this $V$ features Van Hove structure. Note that such a
Hamiltonian occurs, e.g., in the context of spin-boson models at zero
temperature or the scenario addressed by the Weisskopf-Wigner theory, see
\cite{breuer2007}.
\begin{figure}[htb]
\centering
\includegraphics[width=7cm]{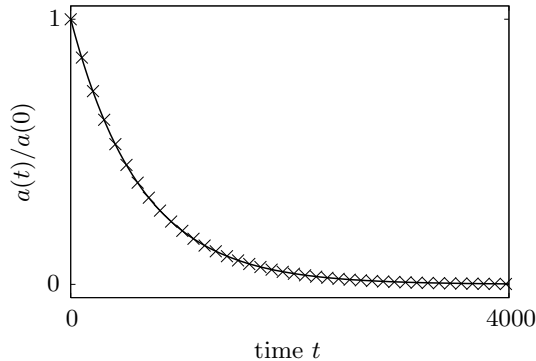}
\caption{Time evolution of the expectation value $a(t)$ for the
interaction of spin-boson type. The numerical result (crosses)
indicates exponential behavior and perfectly agrees with the
theoretical prediction (\ref{eq-astat}) of second order TCL
(continuous curve). System parameters: $n_1 = 1$ (single level),
$n_2 = 2000$ (many levels), $\delta \epsilon = 0.5$, $\lambda =
2.5 \cdot 10^{-4}$.} \label{fig-spinbosWW}
\end{figure}
\\
Fig.~\ref{fig-spinbosWW} shows an almost perfect correspondence
between the numerical solution of the Schr\"odinger equation and
the theoretical prediction (\ref{eq-astat}) which is obtained by
the use of second order TCL. Here, exponential relaxation is
found, although $V$ is not randomly chosen.

\subsection{Sparse Interaction and Localization}

So far, we numerically found exponential decay in accord with the second
order for all considered models that showed the Van Hove structure.
There is, however,
non-exponential behavior for some types of interactions which feature 
the Van Hove property in the sense of (\ref{vanh}) and are in accord with (\ref{eq-crit}). Recall
that those are only 
necessary but not sufficient conditions for the occurrence of
exponential decay.
\\
An example for such a situation is a model with a random but, say, ``sparsely
populated'' interaction. This model is almost identical to the
model with the completely random interaction. The only difference 
is that only $1/10$ of the matrix elements are Gaussian
distributed numbers, all others are zero. The non-zero numbers are
randomly placed. Apparently, this type of interaction fulfills the
Van Hove structure, since the completely random interaction
already does.
\begin{figure}[htb]
\centering
\includegraphics[width=7cm]{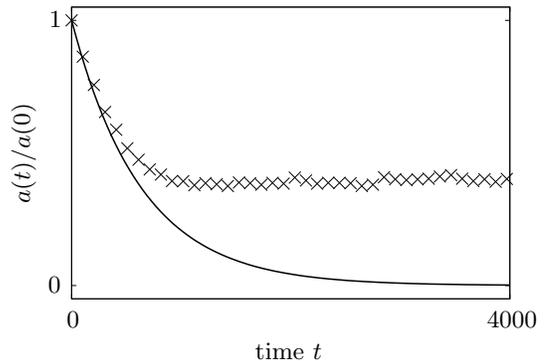}
\caption{Time evolution of the expectation value $a(t)$ for an
interaction with random but ``sparsely populated'' $v_{ij}$. The
theoretical prediction (\ref{eq-astat}) of second order TCL
(continuous curve) deviates from the numerical solution (crosses),
even though the Van Hove structure as well as the conditions
(\ref{eq-crit}) are fulfilled. System parameters: $n = 1000$,
$\delta \epsilon = 0.5$, $\lambda = 2.5 \cdot 10^{-4}$.}
\label{fig-zufThinWW}
\end{figure}
\\
Fig.~\ref{fig-zufThinWW} displays the numerical solution of the
Schr\"odinger equation and the theoretical prediction
(\ref{eq-astat}) of second order TCL. At the beginning there is a
good agreement but then the numerical solution starts to deviate
from a purely exponential decay and finally sticks at a clearly
positive value. The latter non-zero value may be a hint towards
localization effects which also appear, e.g., in the context of
the Anderson model
\cite{anderson1958,abouchacra1973,lee1985,kramer1993}.
And in fact, the sparsely populated interaction takes a form which
is very similar to the Hamiltonian of the, e.g., $3$-dimensional
Anderson model in the chaotic regime.
\\
Apparently, we have to extend the analysis of the fourth order:
There is no exponential behavior by the means of a complete
exponential decay, although $V$ fulfills the Van Hove property.
Recall that the Van Hove criterion has been derived from the
consideration of times $t <\approx\tau_C$ and thus $t = t_1 = t_2 =
t_3<\approx\tau_C$. Hence, we have to reconsider the full time dependence of the fourth
order to produce a feasible estimate for the timescale $t\approx \tau_R$. 
To this end, the integrand $I_1$ is expressed by
\begin{equation}
I_1 = \text{Tr} \{ \, G(t_1, t) \, G(t_2, t_3) \, \} \; ,
\end{equation}
where the hermitian operator $G(t_1, t)$ is again given by
\begin{equation}
G(t, t_1) \equiv [ \, V(t), [ V(t_1), A ] \, ] \; .
\end{equation}
If $I_1(0) \approx C(0)^2$, the diagonal terms dominate at  $t = t_1 =
t_2 =t_3$. Based on this fact, we
carefully assume that $I_1$ is dominated by these terms for other
times as well. Of course, this assumption neglects the larger part
of all terms but leads, as will be demonstrated below, to a criterion
which may be evaluated with limited computational power. (For our
simple example its validity can also be counterchecked by direct numerics.)
However, following this assumption, $I_1$ can be  approximated by
\begin{equation}
I_1 \approx \sum_i G_{ii}(t - t_1) \, G_{ii}(t_2 - t_3) \; ,
\end{equation}
where $G_{ii}(t - t_1)$ are the diagonal matrix elements of $G(t,
t_1)$ in the eigenbasis of $H_0$, namely,
\begin{equation}
G_{ii}(t - t_1) = 2 \sum_{j} (A_{ii} - A_{jj}) \, | V_{ij} |^2 \,
\cos [ \, \omega_{ij} (t - t_1) \, ] \; .
\end{equation}
Furthermore, the correlation function $C(t - t_1)$ can, by the use
of this notation, also be written as
\begin{equation}
C(t - t_1) = \sum_i A_{ii} \, G_{ii}(t - t_1)
\end{equation}
such that $I_2$, the remaining fourth order integrand, can be
expressed as well by
\begin{equation}
I_2 = - \sum_{i, j} A_{ii} \, G_{ii}(t - t_1) \, A_{jj} \, G_{jj}
(t_2 - t_3) \; .
\end{equation}
In order to estimate with reasonable computational effort how $K_4(t)$
compares with $K_2(t)$ another approximation is necessary. Obviously,
the expressions for $I_1,I_2$ are invariant along lines described by
$t_1=$ const., $t_2 = t_3$.
 Thus, as an approximation, we shift the
integration volume from the original region, indicated with
solid lines in  Fig.~\ref{fig-cube}, to a new region, indicated with
dashed lines in  Fig.~\ref{fig-cube}. Obviously, this is a rather rough
estimate but it will turn out to be good enough for our purposes. Now
the coordinate transformation $x = t - t_1$, $y = t_2 - t_3$, $z = t -
t_2$ decouples the integrations within the new integration volume
such that we eventually find for $S(t)\approx K_{4}(t)$ (if $V$ features Van Hove structure)
\begin{equation}
\label{nae}
S(t) = t \, \Big [ \sum_{i} \Gamma_{i}(t)^2 - K_2(t)^2
\Big ] \; ,
\end{equation}
with the time integral $\Gamma_i(t) \equiv \int_{0}^{t} dt' \,
G_{ii}(t')$.
Now (\ref{k4nae})
may eventually be
checked with very low computational power,
based on $S(t)$ from (\ref{nae}).
This adds to (\ref{eq-crit}) and (\ref{vanh}) as a
further manageable criterion for exponential relaxation.
Fig.~\ref{fig-k4app} shows $q(t)$ [based on (\ref{nae})]
for the following interaction types:
the completely random
interaction, the interaction of spin-boson type, and the random
but sparsely populated interaction. 
Fig.~\ref{fig-k4app}
apparently demonstrates that this approximation is able to explain
the breakdown of exponential behavior in the case of a
random, sparsely populated interaction: The fourth order becomes
roughly as large as the second order at a time which agrees with
the deviation between the second order theory and the  numerical
results in Fig.~\ref{fig-zufThinWW}. In both other 
cases $q(t)$ remains sufficiently small, at least until the
relaxation time is reached.
\begin{figure}[htb]
\centering
\includegraphics[width=7cm]{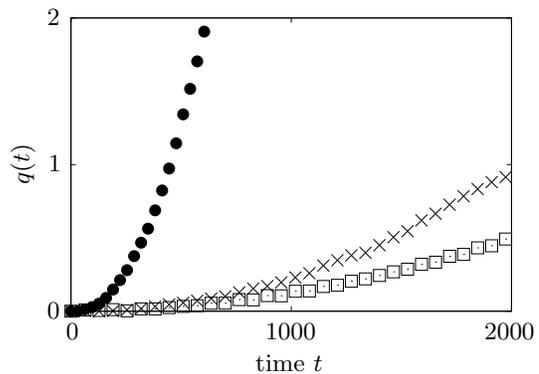}
\caption{
Time evolution of the value $q(t)$ based on (\ref{nae}).
$q(t)$ is numerically calculated
for a completely random interaction (crosses), the interaction of
spin-boson type (squares), and the random but sparsely populated
interaction (circles). Note that $K_2(t)$ is a constant which is
also identical for all three interactions. The system parameters
are chosen according to Fig.~\ref{fig-randWW},
Fig.~\ref{fig-spinbosWW}, and Fig.~\ref{fig-zufThinWW},
respectively.} \label{fig-k4app}
\end{figure}
\\
Obviously, regardless of the interaction type, $K_4(t)$
will eventually dominate  $K_2(t)$ for large enough times (c.f. \cite{khalfin1958}). This, however, does not necessarily
spoil the exponential decay: If $a(t)$ has already decayed almost completely into
equilibrium, even a significant change of the rate $K(t)$ will not
change the overall picture of an exponential decay (, as long as $K(t)$
remains negative). The influence of a large  $K_4(t)$ will only be
visible if it occurs, while  $a(t)$ is still far from equilibrium,
i.e., at times of the order of $\tau_R$. If one now computes the ratio
$q(t)$
for the time $t=\tau_R$, one finds 
\begin{equation}
q(\tau_R)\approx \frac{\sum_i\Gamma_i(\tau_R)^2}{R^2}-1\ ,
\end{equation}
where one has to take (\ref{nae}) and $K_2(\tau_R)=R=1/\tau_R$ into
account. This form has the advantage of being completely independent
of the overall interaction strength $\lambda$. One can hence compute
$q(\tau_R)$, taking $\tau_R$ as a free variable. The region in
which $q(\tau_R)<\approx 1$ then represents the range of different
$\tau_R$ for which exponential decay is possible and to be
expected. The different ``possible'' $\tau_R$ can then be implemented
by tuning $\lambda$ appropriately. Often $q(\tau_R)$ is found to
increase monotonously, essentially like in Fig.~\ref{fig-k4app}. Thus a good
number to characterize a class of models with different relaxation
times (interaction strengths) would be $\tau_{max}$ as the largest time
for which $q(\tau_{max})<\approx 1$ holds true. This then indicates
the largest timescale on which exponential relaxation can still be
expected. We should note here that we intend to use this measure,
$\tau_{max}$, to investigate transport behavior in models of the
Anderson-type in a forthcoming paper.

%
%

\section{Summary and Outlook} \label{sec-summary}

We investigated the dynamics of some expectation values 
for a certain class of closed, finite quantum systems by means of 
the TCL projection operator method. This technique yields a
perturbation expansion for those dynamics. Taking only the second
(leading) order into account, we find that the evolution of these
expectation values may be  
described by a rate equation, i.e., they relax exponentially
if certain criteria are fulfilled. Those criteria,
however, only depend on ``rough'' parameters like overall interaction
strength, bandwidth and density of states but not on, e.g., the phases
of the interaction matrix elements. An adequately computed numerical solution of the 
Schr\"odinger equation is in accord with this leading order result for
random interaction matrices. However, numerics also show that this
accordance breaks down if one considers non-random interactions, even
if the above rough criteria are met. This, of course, indicates that 
higher orders are not negligible, depending on the structure, not only
on the strength of the interaction. Subsequently, we established a
numerically simple estimate for the absolute value of the fourth, i.e., the next higher order, in 
comparison to the second, for short times. From this approach it can
be inferred that the fourth order remains negligible at small times if the interaction features a certain
structure which we define as Van Hove structure according to 
\cite{vanhove1955}. However, numerics indicate that for certain
interaction structures the fourth order may become non-negligible at
larger times, thus spoiling the exponential relaxation, even
if the interaction features Van Hove structure. Hence we suggest one more
criterion (based on (\ref{k4nae}, \ref{nae})) that allows for the detection of such a behavior without
diagonalizing the full system.

Diffusive transport in spatially extended quantum systems may be
viewed as a form of exponential relaxation. Thus we intend to exploit the
various criteria which are suggested in this paper to investigate the
occurrence of diffusion in the Anderson model and/or other solid state
models that do not allow for a full numerical diagonalization.

%
%

\begin{acknowledgments}

We sincerely thank H.-P.~Breuer, M. Michel and M. Kadiro\=glu for their contributions to fruitful
discussions. Financial support by the ``Deutsche
Forschungsgemeinschaft'' is gratefully acknowledged.

\end{acknowledgments}

%
%

\begin{thebibliography}{34}
\expandafter\ifx\csname natexlab\endcsname\relax\def\natexlab#1{#1}\fi
\expandafter\ifx\csname bibnamefont\endcsname\relax
  \def\bibnamefont#1{#1}\fi
\expandafter\ifx\csname bibfnamefont\endcsname\relax
  \def\bibfnamefont#1{#1}\fi
\expandafter\ifx\csname citenamefont\endcsname\relax
  \def\citenamefont#1{#1}\fi
\expandafter\ifx\csname url\endcsname\relax
  \def\url#1{\texttt{#1}}\fi
\expandafter\ifx\csname urlprefix\endcsname\relax\def\urlprefix{URL }\fi
\providecommand{\bibinfo}[2]{#2}
\providecommand{\eprint}[2][]{\url{#2}}

\bibitem[{\citenamefont{{R. Kubo} et~al.}(1991)\citenamefont{{R. Kubo}, {M.
  Toda}, and {N. Hashitsume}}}]{kubo1991}
\bibinfo{author}{\bibnamefont{{R. Kubo}}}, \bibinfo{author}{\bibnamefont{{M.
  Toda}}}, \bibnamefont{and} \bibinfo{author}{\bibnamefont{{N. Hashitsume}}},
  \emph{\bibinfo{title}{Statistical {P}hysics II. {N}onequilibrium
  {S}tatistical {M}echanics}} (\bibinfo{publisher}{Springer, Berlin},
  \bibinfo{year}{1991}).

\bibitem[{\citenamefont{{C. Cercignani}}(1988)}]{cercignani1988}
\bibinfo{author}{\bibnamefont{{C. Cercignani}}}, \emph{\bibinfo{title}{The
  {B}oltzmann {E}quation and its {A}pplications}}
  (\bibinfo{publisher}{Springer, New York}, \bibinfo{year}{1988}).

\bibitem[{\citenamefont{{C. Cohen-Tannoudji} et~al.}(1993)\citenamefont{{C.
  Cohen-Tannoudji}, {B. Diu}, and {F. Lalo\"e}}}]{cohen1993}
\bibinfo{author}{\bibnamefont{{C. Cohen-Tannoudji}}},
  \bibinfo{author}{\bibnamefont{{B. Diu}}}, \bibnamefont{and}
  \bibinfo{author}{\bibnamefont{{F. Lalo\"e}}}, \emph{\bibinfo{title}{Quantum
  {M}echanics}} (\bibinfo{publisher}{Wiley, New York}, \bibinfo{year}{1993}).

\bibitem[{\citenamefont{{M. O. Scully} and {S. Zubairy}}(1997)}]{scully1997}
\bibinfo{author}{\bibnamefont{{M. O. Scully}}} \bibnamefont{and}
  \bibinfo{author}{\bibnamefont{{S. Zubairy}}}, \emph{\bibinfo{title}{Quantum
  {O}ptics}} (\bibinfo{publisher}{Cambridge University Press, Cambridge},
  \bibinfo{year}{1997}).

\bibitem[{\citenamefont{{Van Hove}}(1955)}]{vanhove1955}
\bibinfo{author}{\bibfnamefont{L.}~\bibnamefont{{Van Hove}}},
  \bibinfo{journal}{Physica} \textbf{\bibinfo{volume}{21}},
  \bibinfo{pages}{517} (\bibinfo{year}{1955}).

\bibitem[{\citenamefont{{Van Hove}}(1957)}]{vanhove1957}
\bibinfo{author}{\bibfnamefont{L.}~\bibnamefont{{Van Hove}}},
  \bibinfo{journal}{Physica} \textbf{\bibinfo{volume}{23}},
  \bibinfo{pages}{441} (\bibinfo{year}{1957}).

\bibitem[{\citenamefont{{U. Weiss}}(1999)}]{weiss1999}
\bibinfo{author}{\bibnamefont{{U. Weiss}}}, \emph{\bibinfo{title}{Dissipative
  {Q}uantum {S}ystems}} (\bibinfo{publisher}{World Scientific, Singapore},
  \bibinfo{year}{1999}).

\bibitem[{\citenamefont{{L. P. Kadanoff} and {G. Baym}}(1962)}]{kadanoff1962}
\bibinfo{author}{\bibnamefont{{L. P. Kadanoff}}} \bibnamefont{and}
  \bibinfo{author}{\bibnamefont{{G. Baym}}}, \emph{\bibinfo{title}{Quantum
  {S}tatistical {M}echanics}} (\bibinfo{publisher}{Benjamin, New York},
  \bibinfo{year}{1962}).

\bibitem[{\citenamefont{{A. Fulinski}}(1967)}]{fulinski1967}
\bibinfo{author}{\bibnamefont{{A. Fulinski}}}, \bibinfo{journal}{Phys. Lett. A}
  \textbf{\bibinfo{volume}{25}}, \bibinfo{pages}{13} (\bibinfo{year}{1967}).

\bibitem[{\citenamefont{{A. Fulinski} and {W. J.
  Kramarczyk}}(1968)}]{fulinski1968}
\bibinfo{author}{\bibnamefont{{A. Fulinski}}} \bibnamefont{and}
  \bibinfo{author}{\bibnamefont{{W. J. Kramarczyk}}}, \bibinfo{journal}{Physica
  A} \textbf{\bibinfo{volume}{39}}, \bibinfo{pages}{575}
  (\bibinfo{year}{1968}).

\bibitem[{\citenamefont{{P. H\"anggi} and {H. Thomas}}(1977)}]{hanggi1977}
\bibinfo{author}{\bibnamefont{{P. H\"anggi}}} \bibnamefont{and}
  \bibinfo{author}{\bibnamefont{{H. Thomas}}}, \bibinfo{journal}{Z. Physik B}
  \textbf{\bibinfo{volume}{26}}, \bibinfo{pages}{85} (\bibinfo{year}{1977}).

\bibitem[{\citenamefont{{H. Grabert} et~al.}(1977)\citenamefont{{H. Grabert},
  {P. Talkner}, and {P. H\"anggi}}}]{grabert1977}
\bibinfo{author}{\bibnamefont{{H. Grabert}}}, \bibinfo{author}{\bibnamefont{{P.
  Talkner}}}, \bibnamefont{and} \bibinfo{author}{\bibnamefont{{P. H\"anggi}}},
  \bibinfo{journal}{Z. Physik B} \textbf{\bibinfo{volume}{26}},
  \bibinfo{pages}{389} (\bibinfo{year}{1977}).

\bibitem[{\citenamefont{{S. Chaturvedi} and {F.
  Shibata}}(1979)}]{chaturvedi1979}
\bibinfo{author}{\bibnamefont{{S. Chaturvedi}}} \bibnamefont{and}
  \bibinfo{author}{\bibnamefont{{F. Shibata}}}, \bibinfo{journal}{Z. Phys. B}
  \textbf{\bibinfo{volume}{35}}, \bibinfo{pages}{297} (\bibinfo{year}{1979}).

\bibitem[{\citenamefont{{H. P. Breuer} and {F.
  Petruccione}}(2007)}]{breuer2007}
\bibinfo{author}{\bibnamefont{{H. P. Breuer}}} \bibnamefont{and}
  \bibinfo{author}{\bibnamefont{{F. Petruccione}}},
  \emph{\bibinfo{title}{Theory of {O}pen {Q}uantum {S}ystems}}
  (\bibinfo{publisher}{Oxford University Press, Oxford}, \bibinfo{year}{2007}).

\bibitem[{\citenamefont{{C. D. Blanga} and {M. A.
  Desp\'osito}}(1996)}]{blanga1996}
\bibinfo{author}{\bibnamefont{{C. D. Blanga}}} \bibnamefont{and}
  \bibinfo{author}{\bibnamefont{{M. A. Desp\'osito}}},
  \bibinfo{journal}{Physica A} \textbf{\bibinfo{volume}{227}},
  \bibinfo{pages}{248} (\bibinfo{year}{1996}).

\bibitem[{\citenamefont{{A. A. Budini}}(2006)}]{budini2006}
\bibinfo{author}{\bibnamefont{{A. A. Budini}}}, \bibinfo{journal}{Phys. Rev. A}
  \textbf{\bibinfo{volume}{74}}, \bibinfo{pages}{053815}
  (\bibinfo{year}{2006}).

\bibitem[{\citenamefont{{M. Michel} et~al.}(2005)\citenamefont{{M. Michel}, {G.
  Mahler}, and {J. Gemmer}}}]{michel2005}
\bibinfo{author}{\bibnamefont{{M. Michel}}}, \bibinfo{author}{\bibnamefont{{G.
  Mahler}}}, \bibnamefont{and} \bibinfo{author}{\bibnamefont{{J. Gemmer}}},
  \bibinfo{journal}{Phys. Rev. Lett.} \textbf{\bibinfo{volume}{95}},
  \bibinfo{pages}{180602} (\bibinfo{year}{2005}).

\bibitem[{\citenamefont{{J. Gemmer} et~al.}(2006)\citenamefont{{J. Gemmer}, {R.
  Steinigeweg}, and {M. Michel}}}]{gemmer2006}
\bibinfo{author}{\bibnamefont{{J. Gemmer}}}, \bibinfo{author}{\bibnamefont{{R.
  Steinigeweg}}}, \bibnamefont{and} \bibinfo{author}{\bibnamefont{{M.
  Michel}}}, \bibinfo{journal}{Phys. Rev. B} \textbf{\bibinfo{volume}{73}},
  \bibinfo{pages}{104302} (\bibinfo{year}{2006}).

\bibitem[{\citenamefont{{R. Steinigeweg} et~al.}(2006)\citenamefont{{R.
  Steinigeweg}, {J. Gemmer}, and {M. Michel}}}]{steinigeweg2006}
\bibinfo{author}{\bibnamefont{{R. Steinigeweg}}},
  \bibinfo{author}{\bibnamefont{{J. Gemmer}}}, \bibnamefont{and}
  \bibinfo{author}{\bibnamefont{{M. Michel}}}, \bibinfo{journal}{Europhys.
  Lett.} \textbf{\bibinfo{volume}{75}}, \bibinfo{pages}{406}
  (\bibinfo{year}{2006}).

\bibitem[{\citenamefont{{H. P. Breuer} et~al.}(2006)\citenamefont{{H. P.
  Breuer}, {J. Gemmer}, and {M. Michel}}}]{breuer2006}
\bibinfo{author}{\bibnamefont{{H. P. Breuer}}},
  \bibinfo{author}{\bibnamefont{{J. Gemmer}}}, \bibnamefont{and}
  \bibinfo{author}{\bibnamefont{{M. Michel}}}, \bibinfo{journal}{Phys. Rev. E}
  \textbf{\bibinfo{volume}{73}}, \bibinfo{pages}{016139}
  (\bibinfo{year}{2006}).

\bibitem[{\citenamefont{{R. Steinigeweg} et~al.}(2007)\citenamefont{{R.
  Steinigeweg}, {H. P. Breuer}, and {J. Gemmer}}}]{steinigeweg2007}
\bibinfo{author}{\bibnamefont{{R. Steinigeweg}}},
  \bibinfo{author}{\bibnamefont{{H. P. Breuer}}}, \bibnamefont{and}
  \bibinfo{author}{\bibnamefont{{J. Gemmer}}}, \bibinfo{journal}{Phys. Rev.
  Lett.} \textbf{\bibinfo{volume}{99}}, \bibinfo{pages}{150601}
  (\bibinfo{year}{2007}).

\bibitem[{\citenamefont{{M. Kadiro\=glu} and {J.
  Gemmer}}(2007)}]{kadiroglu2007}
\bibinfo{author}{\bibnamefont{{M. Kadiro\=glu}}} \bibnamefont{and}
  \bibinfo{author}{\bibnamefont{{J. Gemmer}}}, \bibinfo{journal}{Phys. Rev. B}
  \textbf{\bibinfo{volume}{76}}, \bibinfo{pages}{024306}
  (\bibinfo{year}{2007}).

\bibitem[{\citenamefont{{S. Nakajima}}(1958)}]{nakajima1958}
\bibinfo{author}{\bibnamefont{{S. Nakajima}}}, \bibinfo{journal}{Progr. Theor.
  Phys.} \textbf{\bibinfo{volume}{20}}, \bibinfo{pages}{948}
  (\bibinfo{year}{1958}).

\bibitem[{\citenamefont{{R. Zwanzig}}(1960)}]{zwanzig1960}
\bibinfo{author}{\bibnamefont{{R. Zwanzig}}}, \bibinfo{journal}{J. Chem. Phys.}
  \textbf{\bibinfo{volume}{33}}, \bibinfo{pages}{1338} (\bibinfo{year}{1960}).

\bibitem[{\citenamefont{{K. M. Fonseca Romero} et~al.}(2004)\citenamefont{{K.
  M. Fonseca Romero}, {P. Talkner}, and {P. H\"anggi}}}]{romero2004}
\bibinfo{author}{\bibnamefont{{K. M. Fonseca Romero}}},
  \bibinfo{author}{\bibnamefont{{P. Talkner}}}, \bibnamefont{and}
  \bibinfo{author}{\bibnamefont{{P. H\"anggi}}}, \bibinfo{journal}{Phys. Rev.
  A} \textbf{\bibinfo{volume}{69}}, \bibinfo{pages}{052109}
  (\bibinfo{year}{2004}).

\bibitem[{\citenamefont{{M. Esposito} and {P. Gaspard}}(2003)}]{esposito2003}
\bibinfo{author}{\bibnamefont{{M. Esposito}}} \bibnamefont{and}
  \bibinfo{author}{\bibnamefont{{P. Gaspard}}}, \bibinfo{journal}{Phys. Rev. E}
  \textbf{\bibinfo{volume}{68}}, \bibinfo{pages}{066113}
  (\bibinfo{year}{2003}).

\bibitem[{\citenamefont{{J. Gemmer} and {M.
  Michel}}(2006{\natexlab{a}})}]{gemmereur2006}
\bibinfo{author}{\bibnamefont{{J. Gemmer}}} \bibnamefont{and}
  \bibinfo{author}{\bibnamefont{{M. Michel}}}, \bibinfo{journal}{Eur. Phys. J.
  B} \textbf{\bibinfo{volume}{53}}, \bibinfo{pages}{517}
  (\bibinfo{year}{2006}{\natexlab{a}}).

\bibitem[{\citenamefont{{J. Gemmer} and {M.
  Michel}}(2006{\natexlab{b}})}]{gemmereurlett2006}
\bibinfo{author}{\bibnamefont{{J. Gemmer}}} \bibnamefont{and}
  \bibinfo{author}{\bibnamefont{{M. Michel}}}, \bibinfo{journal}{Eurohys.
  Lett.} \textbf{\bibinfo{volume}{73}}, \bibinfo{pages}{1}
  (\bibinfo{year}{2006}{\natexlab{b}}).

\bibitem[{\citenamefont{{J. Gemmer} et~al.}(2004)\citenamefont{{J. Gemmer}, {M.
  Michel}, and {G. Mahler}}}]{gemmer2004}
\bibinfo{author}{\bibnamefont{{J. Gemmer}}}, \bibinfo{author}{\bibnamefont{{M.
  Michel}}}, \bibnamefont{and} \bibinfo{author}{\bibnamefont{{G. Mahler}}},
  \emph{\bibinfo{title}{Quantum {T}hermodynamics}}
  (\bibinfo{publisher}{Springer, New York}, \bibinfo{year}{2004}).

\bibitem[{\citenamefont{Anderson}(1958)}]{anderson1958}
\bibinfo{author}{\bibfnamefont{P.~W.} \bibnamefont{Anderson}},
  \bibinfo{journal}{Phys. Rev.} \textbf{\bibinfo{volume}{109}},
  \bibinfo{pages}{1492} (\bibinfo{year}{1958}).

\bibitem[{\citenamefont{{R. Abou-Chacra} et~al.}(1973)\citenamefont{{R.
  Abou-Chacra}, {D. J. Thouless}, and {P. W. Anderson}}}]{abouchacra1973}
\bibinfo{author}{\bibnamefont{{R. Abou-Chacra}}},
  \bibinfo{author}{\bibnamefont{{D. J. Thouless}}}, \bibnamefont{and}
  \bibinfo{author}{\bibnamefont{{P. W. Anderson}}}, \bibinfo{journal}{J. Phys.
  C} \textbf{\bibinfo{volume}{6}}, \bibinfo{pages}{1734}
  (\bibinfo{year}{1973}).

\bibitem[{\citenamefont{{P. A. Lee} and {T. V. Ramakrishnan}}(1985)}]{lee1985}
\bibinfo{author}{\bibnamefont{{P. A. Lee}}} \bibnamefont{and}
  \bibinfo{author}{\bibnamefont{{T. V. Ramakrishnan}}}, \bibinfo{journal}{Rev.
  Mod. Phys.} \textbf{\bibinfo{volume}{57}}, \bibinfo{pages}{287}
  (\bibinfo{year}{1985}).

\bibitem[{\citenamefont{{B. Kramer} and {A. MacKinnon}}(1993)}]{kramer1993}
\bibinfo{author}{\bibnamefont{{B. Kramer}}} \bibnamefont{and}
  \bibinfo{author}{\bibnamefont{{A. MacKinnon}}}, \bibinfo{journal}{Rep. Progr.
  Phys.} \textbf{\bibinfo{volume}{56}}, \bibinfo{pages}{1469}
  (\bibinfo{year}{1993}).

\bibitem[{\citenamefont{{L. A. Khalfin}}(1958)}]{khalfin1958}
\bibinfo{author}{\bibnamefont{{L. A. Khalfin}}}, \bibinfo{journal}{Sov. Phys.
  JETP} \textbf{\bibinfo{volume}{6}}, \bibinfo{pages}{1053}
  (\bibinfo{year}{1958}).

\end{thebibliography}

\end{document}